\documentclass[aps,prl,reprint,letterpaper,amsmath,amssymb,longbibliography,superscriptaddress]{revtex4-2}
\usepackage{amsmath}
\usepackage{graphicx}

\begin{document}
\title{Stimulated resonant spin amplification reveals millisecond electron spin coherence time of rare-earth ions in solids}
\author{V.~V.~Belykh}
\email[]{belykh@lebedev.ru}
\affiliation{P.~N.~Lebedev Physical Institute of the Russian Academy of Sciences, 119991 Moscow, Russia}
\author{A.~R.~Korotneva}
\affiliation{P.~N.~Lebedev Physical Institute of the Russian Academy of Sciences, 119991 Moscow, Russia}
\author{D.~R.~Yakovlev}
\affiliation{P.~N.~Lebedev Physical Institute of the Russian Academy of Sciences, 119991 Moscow, Russia}
\affiliation{Experimentelle Physik 2, Technische Universit\"{a}t Dortmund, D-44221 Dortmund, Germany}
\affiliation{Ioffe Institute, Russian Academy of Sciences, 194021 St. Petersburg, Russia}
\begin{abstract}
The inhomogeneity of an electron spin ensemble as well as fluctuating environment acting upon individual spins drastically shorten the spin coherence time $T_2$ and hinder coherent spin manipulation. We show that this problem can be solved by the simultaneous application of a radiofrequency (rf) field, which stimulates coherent spin precession decoupled from an inhomogeneous environment, and periodic optical pulses, which amplify this precession. The resulting resonance, taking place when the rf field frequency approaches the laser pulse repetition frequency, has a width determined by the spin coherence time $T_2$ that is free from the effects of inhomogeneity and slow nuclear spin fluctuations. We measure a 50-Hz-narrow electron spin resonance and milliseconds-long $T_2$ for electrons in the ground state of Ce$^{3+}$ ions in the YAG lattice at low temperatures, while the inhomogeneous spin dephasing time $T_2^*$ is only 25~ns. This study paves the way to coherent optical manipulation in spin systems decoupled from their inhomogeneous environment. 
\end{abstract}
\maketitle

Periodic optical orientation of an electron spin ensemble in a constant magnetic field $\mathbf{B}$ can lead to the enhancement of the total spin polarization if optical pulses come in phase with the precessing spins. This effect called resonant spin amplification (RSA) takes place when the Larmor frequency of spin precession $f_\text{L}$ in the magnetic field is a multiple of the optical pulse repetition frequency $f_\text{o}$: $f_\text{L} = m f_\text{o}$, $m = 0,1,2...$ \cite{Kikkawa1998,Glazov2008,Yugova2012,Saeed2018}. Similarly, application of an oscillating radio frequency (rf) field to an ensemble of electron spins leads to electron spin resonance (ESR) \cite{Blume1958,Gordon1958,Schweiger2001} when the rf field frequency $f_\text{rf}$ is equal to the Larmor frequency: $f_\text{rf} = f_\text{L}$.  
The resonance frequency and the width of either RSA or ESR resonances give the average $g$ factor and the inhomogeneous spin dephasing time $T_2^*$ for the spin ensemble. The  time $T_2^*$ in systems with localized electrons is often dominated by the dephasing of the spin ensemble caused by a spread in the Larmor frequencies of different electrons. It is much shorter than the spin coherence time $T_2$ of individual spin. Measurement of $T_2$ is demanding, it requires addressing individual spins \cite{Siyushev2014,Bechtold2016} and/or implementing the spin echo technique \cite{Hahn1950,Azamat2017,Bechtold2015,DeGreve2011}. The time $T_2$ measured in these sophisticated experiments is often limited by the time-fluctuating environment, such as nuclear effective fields, varying on a timescale shorter than $T_2$. To resolve this issue, an electron spin is dynamically decoupled from the environment by applying a sequence of pulses flipping the spin state \cite{Carr1954,Meiboom1958,Viola1998,Siyushev2014}, which dramatically increases $T_2$ but further complicates the experiments. 

In this study we show that the simultaneous application of a periodic optical pumping and a continuous-wave rf magnetic field to an inhomogeneous electron spin ensemble results in a sharp resonance at $f_\text{rf} = mf_\text{o}$. The width of the resonance gives the spin coherence time $T_2$ free from the effects of ensemble inhomogeneity and fluctuating nuclear environment. This is in contrast to the combined RSA-ESR resonance in homogeneous system, where ESR generally suppresses RSA \cite{Belykh2020}.
 
\begin{figure}
\includegraphics[width=\columnwidth]{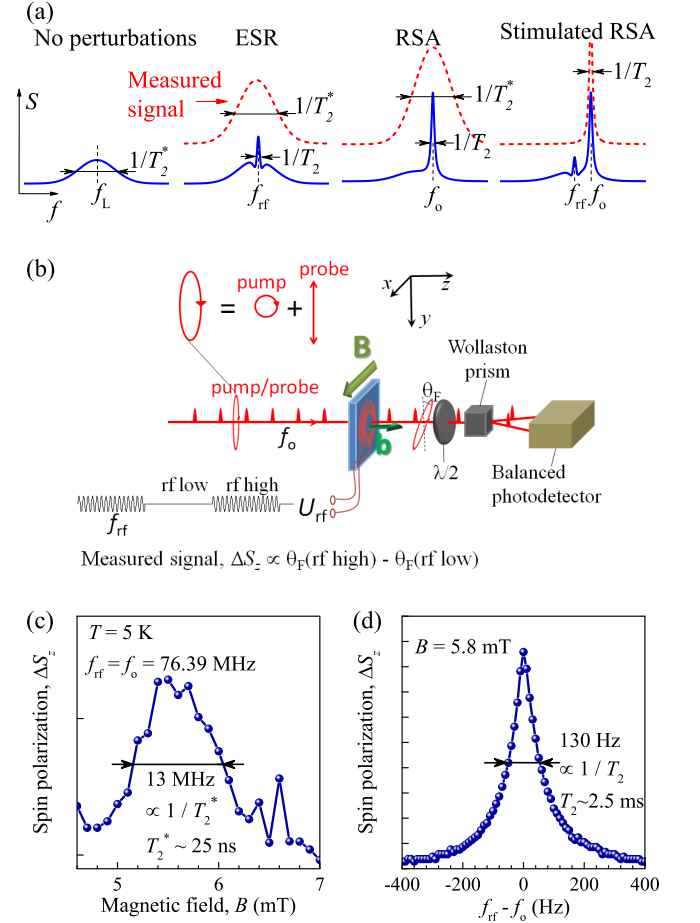}
\caption{(a) Schematic illustration of stimulated RSA. Solid lines show spin polarization frequency distributions in a magnetic field. Dashed lines show the measured signal profiles when $f_\text{rf}$ or the magnetic field are scanned resulting in ESR, RSA, or SRSA spectra. 
(b) Scheme of the SRSA experiment. (c) Spin polarization of Ce$^{3+}$ ions in YAG as a function of the magnetic field for $f_\text{rf} = f_\text{o}$. The width of the peak gives inhomogeneous spin dephasing time $T_2^*$. (d) Spin polarization as a function of the rf field frequency offset with respect to the laser pulse repetition frequency for $B = 5.8$~mT. The width of the peak gives the spin coherence time $T_2$. In (c) and (d), the laser power is $P = 0.5$~mW and the temperature is $T = 5$~K.}
\label{fig:pics}
\end{figure}

This principle is illustrated in Fig.~\ref{fig:pics}(a). Optical pulses applied to an inhomogeneous electron ensemble create and amplify spin polarization for a small subensemble with Larmor frequencies $f_\text{o}-1/T_2 \lesssim f_\text{L} \lesssim f_\text{o}+1/T_2$. However, when the permanent magnetic field is scanned, the fixed value of $f_\text{o}$ goes consecutively through the values of $f_\text{L}$ for the entire spin ensemble resulting in a broad RSA curve with a width of $\sim 1/T_2^*$. When a rf field is applied, it synchronizes electron spins and decouples them from the inhomogeneous environment forcing to precess at the common frequency $f_\text{rf}$ \cite{Belykh2019}. This results in a narrow peak in the spin frequency distribution with a width of $\sim 1/T_2$ \cite{Poshakinskiy2020}. Scanning of $f_\text{rf}$ over the Larmor frequencies of the ensemble results, nevertheless, in a broad ESR curve with a width of $\sim 1/T_2^*$. When both optical pulses and rf field are applied to the spin ensemble and $f_\text{rf} \approx f_\text{o}$, the rf field stimulates RSA by providing homogenized spin subensemble which spin polarization is optically amplified and can be detected experimentally. Therefore, we refer this principle as stimulated resonant spin amplification (SRSA). 
When $f_\text{rf}$ is scanned across  $f_\text{o}$ the width of the resonance is determined by the coherence time $T_2$ of electron spins decoupled from inhomogeneous environment. 
The experiments can be performed at any magnetic field provided $f_\text{L} \approx f_\text{rf}  \approx m f_\text{o}$.  
We demonstrate this principle on the ensemble of rare-earth Ce$^{3+}$ ions in the YAG crystal, where we measure $T_2 = 9$~ms at liquid-helium temperatures. This is the largest value reported so far for Ce$^{3+}$:YAG, while $T_2^*$ is limited to about 25~ns.

The sample under study is a 0.5-mm-thick Ce$^{3+}$:YAG crystal with a Ce$^{3+}$
ion concentration of 0.5 at. \%. The scheme of the experiment shown in Fig.~\ref{fig:pics}(b) is rather simple. The sample is placed in a variable temperature ($5-300$~K) He-flow cryostat.  Using a permanent magnet placed outside the cryostat at a controllable distance from the sample, a constant magnetic field $\mathbf{B}$ up to 20~mT is applied along the $x$ axis, which is perpendicular to the direction of light propagation ($z$ axis) and to the sample normal (Voigt geometry). The optical spin pumping and probing are performed by the same laser beam with elliptical initial polarization. The circular and linear components of the elliptically polarized beam can serve as the simultaneous pump and probe, respectively, for the electron spin~\cite{Belykh2020,Ryzhov2016}. Rigorous analysis of the effect of the laser beam ellipticity on the measured signal is given in the Supplemental Material. We use a pulsed Ti:Sapphire laser operating at a wavelength of 888~nm that is frequency doubled with a BBO crystal to obtain a wavelength of 444 nm. The laser generates a train of 2-ps-long optical pulses with a repetition frequency $f_\text{o} = 76.39$~MHz. We measure the spin polarization via the Faraday rotation of the linear polarization component of the laser beam transmitted through the sample. It is analyzed using a Wollaston prism, splitting the beam into two orthogonally polarized beams of approximately equal intensities that are further registered by a balanced photodetector. 

The rf magnetic field is applied along the sample normal ($z$ axis) using a small coil ($1$~mm-inner and $1.5$~mm-outer diameter) near the sample surface. Current through the coil is driven by a function generator, which creates a sinusoidal voltage with a frequency $f_\text{rf}$ up to 150~MHz and an amplitude $U_\text{rf}$ up to 10~V. The generator output is modulated at a frequency of 5~kHz for synchronous detection with a lock-in amplifier. Thus, the measured signal is proportional to the difference between the Faraday rotation values for the high and low levels of the rf field, which is in turn proportional to the corresponding difference $\Delta S_z$ in the $z$-components of the spin polarizations~\cite{Belykh2020}. 

The energy level structure of the Ce$^{3+}$ ion and the scheme its of optical orientation can be found in Refs.~\cite{Kolesov2013,Siyushev2014,Azamat2017,Liang2017}. This ion has one unpaired electron in the 4f level, which can be excited optically to the 5d level via the phonon-assisted absorption. 
Circularly polarized light excites electrons with a certain spin (spin-down in the case of $\sigma^+$ polarization) which is flipped in the course of excitation. Meanwhile, upon their relaxation back to the ground 4f level, electrons may end up in spin-down or spin-up state with equal probability. In this way the electrons occupying the ground 4f level for the ensemble of Ce$^{3+}$ ions become preferentially spin-up polarized under $\sigma^+$ excitation. For periodic circularly-polarized optical pumping the spin polarization is enhanced if the Larmor frequency satisfies the RSA condition, i.e., $f_\text{L} = m f_\text{o}$. RSA for Ce$^{3+}$ in YAG was already observed in the two-pulse experiment \cite{Azamat2017}.

First, we measure the effect of a rf magnetic field on the optically amplified spin polarization by scanning the permanent magnetic field $B$ with the rf field frequency fixed  at $f_\text{rf} = f_\text{o}$, as shown in Fig.~\ref{fig:pics}(c). Only the spins with the Larmor frequency $f_\text{L} = |g| \mu_\text{B} B/2 \pi \hbar = f_\text{rf} = f_\text{o}$ are optically amplified and addressed by the rf field. Thus, the broad peak in Fig.~\ref{fig:pics}(c) at $B = 5.5$~mT corresponds to $|g| \approx 1.0$. Its full width at half maximum (FWHM) $\delta B \approx 1$~mT gives the spread in the Larmor frequencies $\delta f_\text{L} = 13$~MHz related to a spread in $g$ factors and nuclear fields within the spin ensemble. This spread leads to the dephasing of the spin ensemble with $T_2^* = 1 / \pi \delta f_\text{L} \approx 25$~ns being in agreement with Ref.~\cite{Azamat2017}.  

Second, we fix the magnetic field at 5.8~mT, so that $f_\text{o}$ is within the broad distribution of $f_\text{L}$, and scan the rf field frequency $f_\text{rf}$. The dependence of the spin polarization $\Delta S_z$ on $f_\text{rf}$ [Fig.~\ref{fig:pics}(d)] shows an extremely sharp peak at $f_\text{rf} = f_\text{o}$ with FWHM $\delta f_\text{rf} \approx 130$~Hz. This peak can be interpreted as rf-stimulated RSA and its width gives the spin coherence time $T_2 = 1 / \pi \delta f_\text{rf} \approx 2.5$~ms. 

\begin{figure*}
\includegraphics[width=2\columnwidth]{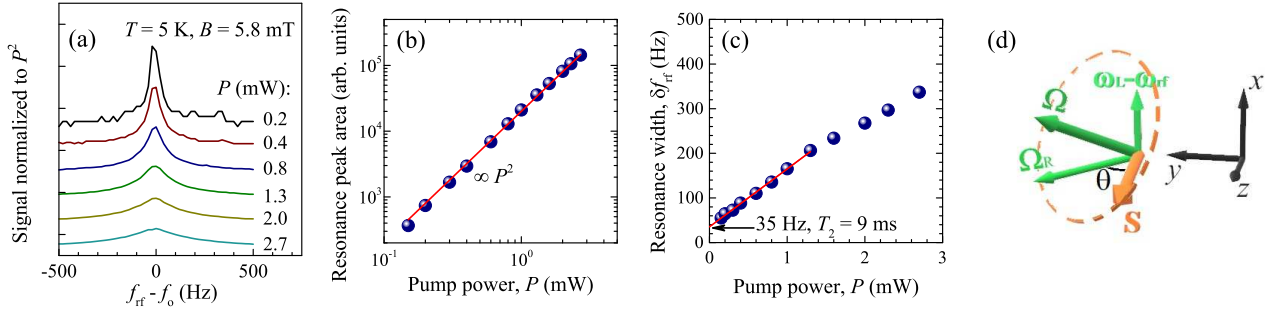}
\caption{(a) Faraday rotation signal (difference of the intensities registered by the balanced photodetector) as a function of the rf field frequency offset with respect to the laser pulse repetition frequency (SRSA spectra) for different laser powers $P$. The spectra are normalized to $P^2$. (b) Laser power dependence of the resonance peak area. The solid line shows the quadratic dependence. (c) Laser power dependence of the resonance peak width. The solid line shows the linear dependence. In (a)-(c) $B = 5.8$~mT, and $T = 5$~K. (d) Schematics of spin precession in the rotating reference frame. The optically excited spin $\mathbf{S}$ precesses at a frequency $\boldsymbol{\Omega} = \boldsymbol{\Omega}_\text{R} + \boldsymbol{\omega}_\text{L}-\boldsymbol{\omega}_\text{rf}$.}
\label{fig:PDep}
\end{figure*}

The magnitude (area) and width of the SRSA peak strongly depend on the laser power $P$, see Figs.~\ref{fig:PDep}(a)-(c). Its magnitude is proportional to $P^2$, which gives clear evidence that the laser beam not only probes spin polarization, but simultaneously pumps it. Indeed, the signal registered by the balanced photodetector shown in Figs.~\ref{fig:PDep}(a) is proportional to $P$ times the Faraday rotation angle. The later is proportional to the spin polarization, which is also proportional to $P$ (see also the Supplemental Material). Note, the signal at the peak maximum increases slower than $P^2$ due to the broadening [Fig.~\ref{fig:PDep}(a)] i.e. decrease of $T_2$. Peak FWHM $\delta f_\text{rf}$ linearly increases with $P$ [Fig.~\ref{fig:PDep}(c)], which may be related to the fact that, apart from creating spin polarization, the pump also disturbs the coherent precession of the spin polarization oriented by previous pump pulses. In the limit of $P \rightarrow 0$ we get $\delta f_\text{rf} = 35$~Hz and $T_2 = 9$~ms, corresponding to the unperturbed system. Note that the longest $T_2$ time reported for Ce$^{3+}$:YAG so far was 2~ms. It was measured for single ions with the application of a decoupling rf protocol \cite{Siyushev2014}.

\begin{figure}
\includegraphics[width=1\columnwidth]{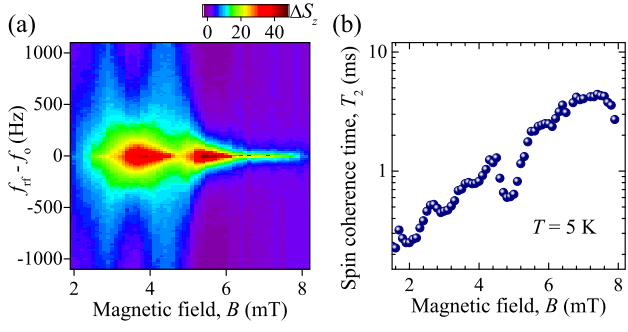}
\caption{(a) Spin polarization as a function of the magnetic field and the rf field frequency offset with respect to the laser pulse repetition frequency. (b) Magnetic field dependence of the spin coherence time $T_2$. In (a),(b) $P = 0.5$~mW, and $T = 5$~K.}
\label{fig:BDep}
\end{figure}

The dependence of the spin polarization on both $f_\text{rf}$ and the magnetic field is shown in Fig.~\ref{fig:BDep}(a). The peaks at different fields correspond to the set of different $g$-factors. At increased laser power ($P = 1$~mW) the peaks at $B = 2.2, 3.3$ and $5.5$~mT  become better resolved despite a decrease in $T_2$ (Fig.~S3 in the Supplemental Material). They correspond to $|g| = 2.5, 1.7$, and $1.0$. For an electron at the 4f level of a Ce$^{3+}$ ion the $g$ tensor is highly anisotropic. Ce$^{3+}$ ions can occupy $c$ sites in the YAG matrix with six possible orientations of the $g$ tensor. Correspondingly, six different $g$ factors ranging from 0.9 to 2.7 can be observed for a given orientation of the magnetic field \cite{Lewis1966,Azamat2017}. Note, that peaks corresponding to close values of the $g$ factor are not resolved in the magnetic field dependence at low fields.  
The elongated high-field tails of the peaks characterized by increased values of $T_2$ may be attributed to the buildup of the nuclear spin polarization, which changes the effective magnetic field acting upon an electron spin in analogy to \cite{Greilich2007,Zhukov2018a,Evers2021}.
The dependence of $T_2$ on the magnetic field is shown in Fig.~\ref{fig:BDep}(b). Note that here an increase in the magnetic field corresponds to a decrease in the $g$ factor, while the Larmor frequency of the electrons for which $T_2$ is determined stays in the vicinity of $f_\text{o} = 76.36$~MHz. The dependence features a number of sharp dropdowns, but generally $T_2$ increases by more than one order of magnitude as the field increases from 2 to 6 mT and then saturates at about 6~ms. The initial increase in $T_2$ with $B$ may be related to the overcoming of the nuclear field fluctuations, having an amplitude of few mT, in an external magnetic field \cite{Merkulov2002,Liang2017,Azamat2017}.

\begin{figure}
\includegraphics[width=1\columnwidth]{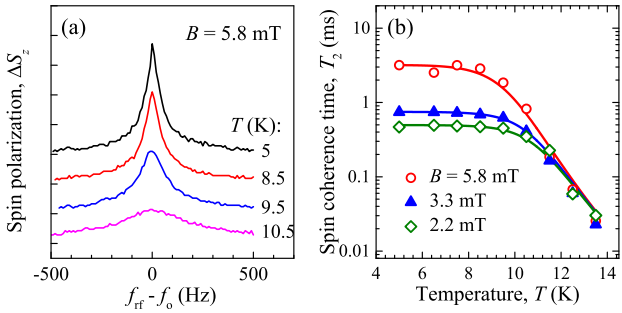}
\caption{(a) SRSA spectra at different temperatures for $B = 5.8$~mT. (b) Temperature dependence of the spin coherence time $T_2$ for different magnetic fields. The solid lines show fits to the experimental data with Eq.~\eqref{eq:LO}. In (a),(b) $P = 0.5$~mW.}
\label{fig:TDep}
\end{figure}

An increase in the temperature from 5 to 13~K expectedly results in the broadening of the SRSA spectra [Fig.~\ref{fig:TDep}(a)]. The temperature dependencies of $T_2$ are shown in Fig.~\ref{fig:TDep}(b) for different magnetic fields. 
The observed decrease in $T_2$ with $T$ can be described by the two-phonon Raman process ($T^9$) modified by the presence of a longitudinal optical (LO) phonon mode:
\begin{equation}
1 / T_2 (T) = 1/T_2(T=0) + A T^9 + C \exp(-E_\text{a}/k_\text{B}T).
\label{eq:LO}
\end{equation} 
All curves are fitted with the parameters $A = 4 \times 10^{-7}$~s$^{-1}$K$^{-9}$, $C = 1.4 \times 10^{10}$~s$^{-1}$ and $E_\text{a} = 125$~cm$^{-1}$, corresponding to the energy  of the LO phonon in YAG \cite{Brog1966}. Similar parameters were used in Ref.~\cite{Azamat2017} to fit the temperature dependence of $T_1$ in the same sample. This confirms that the observed resonances with different $g$ factors have the same origin (4f state of the Ce$^{3+}$ ion) and at $T \gtrsim 10$~K time $T_2$ is limited by inelastic spin relaxation similarly to $T_1$. 


We observe remarkably long times $T_2$ in Ce$^{3+}$:YAG, which have the comparable millisecond-range values and a similar temperature dependence to $T_1$. It is much longer than $T_2^* \approx 25$~ns measured with RSA or ESR. Moreover, it is three orders of magnitude longer than $T_2 = 5$~$\mu$s measured in this system with the spin echo technique 
\cite{Azamat2017}. The conventional spin echo technique makes it possible to overcome time-independent inhomogeneity, such as a spread in $g$ factors and frozen fluctuations of the nuclear field. However, 
$T_2$ measured with the spin echo is limited by the slow variation of the nuclear field between the echo pulses. Thus, our method allows one not only to overcome the inhomogeneity of the system, but also get rid of the contributions from slowly varying nuclear spin fluctuations thanks to continuous driving of the spin ensemble by the rf field. This can be understood by considering the classical Bloch-equation picture of spin precession in a magnetic field \cite{Bloch1946}.

The spin polarization $\mathbf{S}^n$ created by the $n$-th optical pulse precesses about the permanent magnetic field $\mathbf{B}$ with the Larmor frequency $\boldsymbol{\omega}_\text{L} = g\mu_\text{B}\mathbf{B}/\hbar$ ($\boldsymbol{\omega}_\text{L}$ is assumed to be fixed so far, averaging over the spin ensemble will be done later)    
and decays exponentially with the characteristic time $T_2$. This follows from the Bloch equation with omitted equilibrium spin polarization, which is small compared to the optically created one. The additional oscillating magnetic field $\mathbf{b}(t)$ applied by the rf coil can be represented as a sum of the two fields with amplitude $b/2$ rotating in the $yz$ plane in opposite directions with the frequencies $\boldsymbol{\omega}_\text{rf}$ and $-\boldsymbol{\omega}_\text{rf}$ directed parallel and antiparallel to  $\boldsymbol{\omega}_\text{L}$, respectively. The counter-rotating term is strongly out of resonance and can be neglected \cite{Abragam1961}. The action of co-rotating term becomes evident in the reference frame rotating with the frequency $\boldsymbol{\omega}_\text{rf}$, where the field $\mathbf{b}/2$ is constant [Fig.~\ref{fig:PDep}(d)]. Here, the spin precesses with the frequency $\boldsymbol{\Omega} = \boldsymbol{\Omega}_\text{R} + \boldsymbol{\omega}_\text{L}-\boldsymbol{\omega}_\text{rf}$, where $\boldsymbol{\Omega}_\text{R} = g \mu_\text{B} \mathbf{b} /2\hbar$ is the Rabi frequency:
\begin{multline}
\mathbf{S}^n(t) = \{[\mathbf{S}^n(0)-(\mathbf{S}^n(0)\mathbf{e})\mathbf{e}]\cos(\Omega t)+\mathbf{e} \times \mathbf{S}^n(0)\sin(\Omega t) \\
+ (\mathbf{S}^n(0)\mathbf{e})\mathbf{e}\}\exp(-t/T_2),
\label{eq:spin}
\end{multline} 
where $\mathbf{e} = \boldsymbol{\Omega}/\Omega$, $\Omega = \sqrt{\Omega_\text{R}^2+(\omega_\text{L}-\omega_\text{rf})^2}$. The first two terms in Eq.~\eqref{eq:spin} represent the precession of the spin component perpendicular to $\boldsymbol{\Omega}$, while the third term corresponds to the spin component along $\boldsymbol{\Omega}$. It is the third term that represents stimulated precession with the frequency $\omega_\text{rf}$ in the laboratory reference frame and it is weakly sensitive to the variations in $\omega_\text{L}$. Note, the experiments are performed at the rf field amplitude of about 1~mT, corresponding to the Rabi frequency $\Omega_\text{R}/2\pi$ in the MHz range. It is much higher than the expected rate of the nuclear spin fluctuations variation allowing for their efficient suppression.  The experimental dependence of the SRSA signal on the rf field amplitude is presented in the Supplemental Material. Averaging Eq.~\eqref{eq:spin} over inhomogeneous distribution of $\omega_\text{L}$, contributed by the spread of $g$ factors and frozen nuclear field fluctuations, as well as over the slow variations of $\boldsymbol{\omega}_\text{L}$ along $x$ axis (on a time scale larger than $1/\Omega$), zeroes out the two first terms and modifies the third term:
\begin{multline}
<\mathbf{S}^n(t)> = \boldsymbol{\Omega}_\text{R}\Omega_\text{R}<\frac{1}{\Omega^2}>\Delta S \cos(\theta_n)\exp(-t/T_2),
\label{eq:spinav}
\end{multline} 
where $\Delta S = |\mathbf{S}^n(0)|$, $\theta_n = \theta_0 + n \omega_\text{rf} T_\text{o}$ is the angle between $\boldsymbol{\Omega}_\text{R}$ and $\mathbf{S}^n(0)$, $T_\text{o} = 1/f_\text{o}$ is the optical pulse repetition period, and we leave only the spin component in the $yz$ plane, which is measured in the experiment. 
Summing up spin polarization created by all pulses that already arrived at time moments $t_n = n T_\text{o}$ in the past, taking the component along $\mathbf{S}^0(0)$, i.e. along the pump/probe beam, and averaging over $\theta_0$ since the laser train and rf field are not synchronized we obtain:
\begin{multline}
\frac{<S_z>}{\Delta S} = <\frac{\Omega_\text{R}^2}{4\Omega^2}>\\
\times\left[1+\frac{\sinh(T_\text{o}/T_2)}{\cosh(T_\text{o}/T_2)-\cos(\omega_\text{rf}T_\text{o})}\right],
\end{multline} 
This formula resembles the classical RSA expression \cite{Glazov2008,Yugova2012,Azamat2017} with $T_2^*$ replaced with inhomogeneity-free $T_2$. In the vicinity of the RSA peak taking into account that $T_2 \gg T_\text{o}$, we get
\begin{equation}
\frac{<S_z>}{\Delta S}
\approx <\frac{\Omega_\text{R}^2}{2\Omega^2}>\frac{T_2}{T_\text{o}}\times\frac{1}{1+T_2^2(\omega_\text{rf}-m\omega_\text{o})^2}.
\label{eq:spinres}
\end{equation} 
Therefore, the SRSA spectrum can be described by a Lorentzian with a FWHM $\delta f_\text{rf} = \delta \omega_\text{rf} /2\pi = 1/\pi T_2$.
The same expression can be obtained by taking advantage of the theory of the combined RSA-ESR resonance for a homogeneous system developed in Ref.~\cite{Belykh2020} by averaging the spin polarization over the distribution of Larmor frequencies. A more elaborated analysis is needed to show that the measured $T_2$ is robust with respect to the slowly varying fluctuations of $\boldsymbol{\omega}_\text{L}$ in the direction transverse to $\boldsymbol{\omega}_\text{rf}$.


\begin{acknowledgments}
We are grateful to M.~Bayer, M.M.~Glazov, M.~L.~Skorikov, D.~S.~Smirnov and D.~N.~Sob'yanin for fruitful discussions and to D.~H.~Feng for providing the sample. The work was supported by the Government of the Russian Federation, Contract No. 075-15-2021-598 at the P.N. Lebedev Physical Institute (experimental studies), and by the Russian Science Foundation through grant No. 18-72-10073 (development of the technique and modeling). 
\end{acknowledgments}

\newpage

\clearpage

\begin{widetext}
\begin{center}
	\textbf{\large Supplemental material: Stimulated resonant spin amplification reveals millisecond spin coherence time of rare-earth ions in solids}
\end{center}
\end{widetext}

\setcounter{equation}{0}
\setcounter{figure}{0}
\setcounter{table}{0}
\setcounter{section}{0}
\setcounter{page}{1}
\renewcommand{\theequation}{S\arabic{equation}}
\renewcommand{\thefigure}{S\arabic{figure}}
\renewcommand{\bibnumfmt}[1]{[S#1]}
\renewcommand{\citenumfont}[1]{S#1}
\renewcommand{\thetable}{S\arabic{table}}
\renewcommand{\thesection}{S\arabic{section}}

\section{Dependence of the signal on the laser polarization}
\begin{figure}[!h]
\includegraphics[width=0.8\columnwidth]{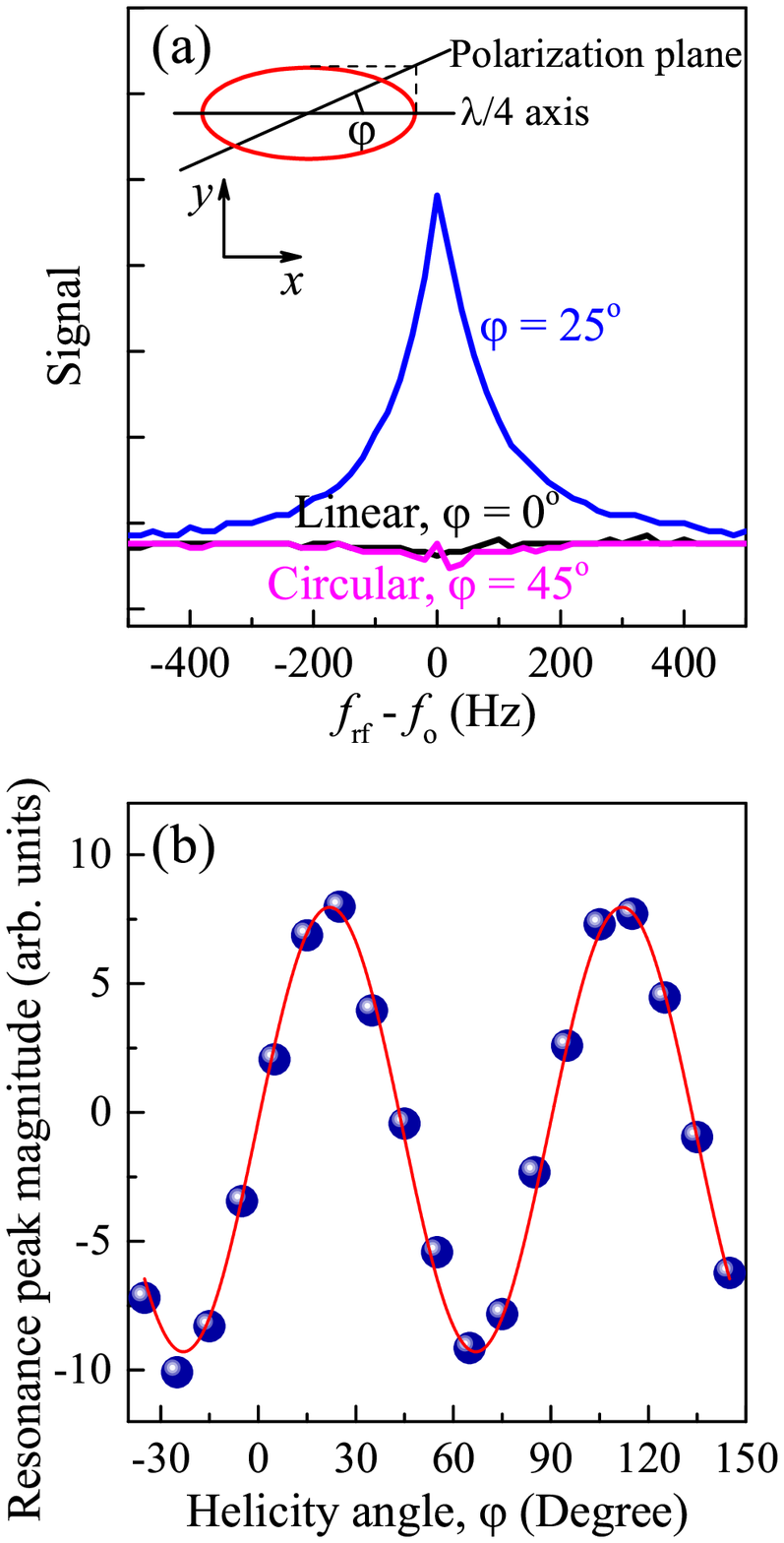}
\caption{(a) Faraday rotation signal (difference of the intensities registered by the balanced photodetector) as a function of the rf field frequency offset with respect to the laser pulse repetition frequency (SRSA spectra) for the different helicities of the laser beam, parametrized by the angle $\varphi$ as shown in the inset. (b) Magnitude of the SRSA resonance peak as a function of the helicity angle $\varphi$. The solid line shows fit with $\sin(4\varphi)$. In (a),(b) the magnetic field is $B = 5.8$~mT, the laser power is $P = 0.5$~mW, and the temperature is $T = 5$~K.}
\label{fig:ADep}
\end{figure}

In our experimental technique it is crucial that the same laser beam has two functions: it excites the spin polarization and probes it through the Faraday rotation. This is evidenced by the quadratic dependence of the signal on the laser power (Fig.~2 in the main text). Here we further prove this fact by analyzing the dependence of the signal on the helicity of the laser beam. The helicity is controlled using a linear polarizer after which a $\lambda/4$ waveplate is installed, so that we are able to control the angle $\varphi$ between the linear polarization plane and the waveplate fast axis. Thus, the axes of the resulting polarization ellipse are $E_0 \cos\varphi$ and $E_0 \sin\varphi$ [inset in Fig.~\ref{fig:ADep}], where $E_0$ is the amplitude of the wave transmitted through the linear polarizer. The field vector of this elliptical polarization incident on the sample is   
\begin{equation}
\mathbf{E_\text{in}} = 	E_0 \begin{pmatrix}
				\cos\varphi\cos(\omega t)\\
				\sin\varphi\sin(\omega t)
				\end{pmatrix},
\end{equation}
where $\omega$ is the light frequency, and the $x$ and $y$ axes are directed along the waveplate axes.
It can be represented a sum of the two orthogonal circular polarizations:
\begin{multline}
\mathbf{E} = 	E_0 \frac{\cos\varphi+\sin\varphi}{2} \begin{pmatrix}
				\cos(\omega t)\\
				\sin(\omega t)
				\end{pmatrix}+\\
				E_0 \frac{\cos\varphi-\sin\varphi}{2} \begin{pmatrix}
				\cos(\omega t)\\
				-\sin(\omega t)
				\end{pmatrix}.
\end{multline}
The excited spin population is proportional to the difference in the intensities of these two components $S_z \propto I_+ - I_- = I_0 [(\cos\varphi+\sin\varphi)^2/4-(\cos\varphi-\sin\varphi)^2/4] = I_0 \sin(2\varphi)/2\large$, where $I_0$ is the intensity of the wave transmitted through the linear polarizer.
The spin polarization in the sample induces the Faraday rotation of the transmitted light polarization which manifests itself as a small rotation of the ellipse axes or as a small additional phase $\delta \phi$ acquired by one circular polarization ($\sigma^+$) with respect to the other one ($\sigma^-$). The field vector transmitted through the sample is
\begin{multline}
\mathbf{E}_\text{out} = E_1 \frac{\cos\varphi+\sin\varphi}{2} \begin{pmatrix}
				\cos(\omega t+\delta\phi)\\
				\sin(\omega t+\delta\phi)
				\end{pmatrix}+\\
				E_1 \frac{\cos\varphi-\sin\varphi}{2} \begin{pmatrix}
				\cos(\omega t)\\
				-\sin(\omega t)
				\end{pmatrix},
\end{multline}
where $E_1 = \sqrt{\mathcal{T}}E_0$, and $\mathcal{T}$ is the sample transmission coefficient (by the intensity). 
The output light is then split by the two beams having the orthogonal linear polarizations and equal intensities in the absence of the Faraday rotation using a Wollaston prism. This can be done if the axes of the Wollaston prism are inclined 45$^\circ$ with respect to the axes of the polarization ellipse i.e. to the $x$ and $y$ axes. Taking projections of the field $\mathbf{E}_\text{out}$ on the Wollaston prism axes, calculating the corresponding intensities by squaring field amplitudes, and taking their difference we obtain the signal measured by the balanced photodetector $\delta I = \mathcal{T} I_0 \cos(2\varphi) \delta \phi/2$. The phase responsible for the Faraday rotation $\delta \phi \propto S_z \propto I_0 \sin(2\varphi)/2$ as it was shown above. Finally the signal detected by the balanced photodetector:
\begin{equation}
\delta I \propto \frac{1}{4}\mathcal{T} I_0^2\sin(2\varphi)\cos(2\varphi)= \frac{1}{8}\mathcal{T} I_0^2\sin(4\varphi).
\label{eq:pol}
\end{equation} 

In this way if $\varphi = 0^\circ$, the linearly polarized light does not excite spin polarization and produces no signal, if $\varphi = 45^\circ$, the circularly polarized light does not probe spin polarization and produces no signal, while at an intermediate angle, e.g. $\varphi = 25^\circ$, light both excites and probes spin polarization [Fig.~\ref{fig:ADep}(a)]. The dependence of the signal magnitude measured as the area of the stimulated resonant spin amplification (SRSA) resonance peak (which sign follows the sign of the Faraday rotation) on the incident light ellipticity is shown in Fig.~\ref{fig:ADep}(b). It indeed shows $\sin(4\varphi)$ dependence. Furthermore, one can see from Eq.~\eqref{eq:pol} that the registered signal is proportional to the squared laser beam intensity $I_0^2$ (or power $P^2$) in agreement with Fig.~2 in the main article. We also note that the resonance width, and thus $T_2$, weakly depends on the laser helicity and is affected by the total laser power. 

\begin{figure}
\includegraphics[width=0.9\columnwidth]{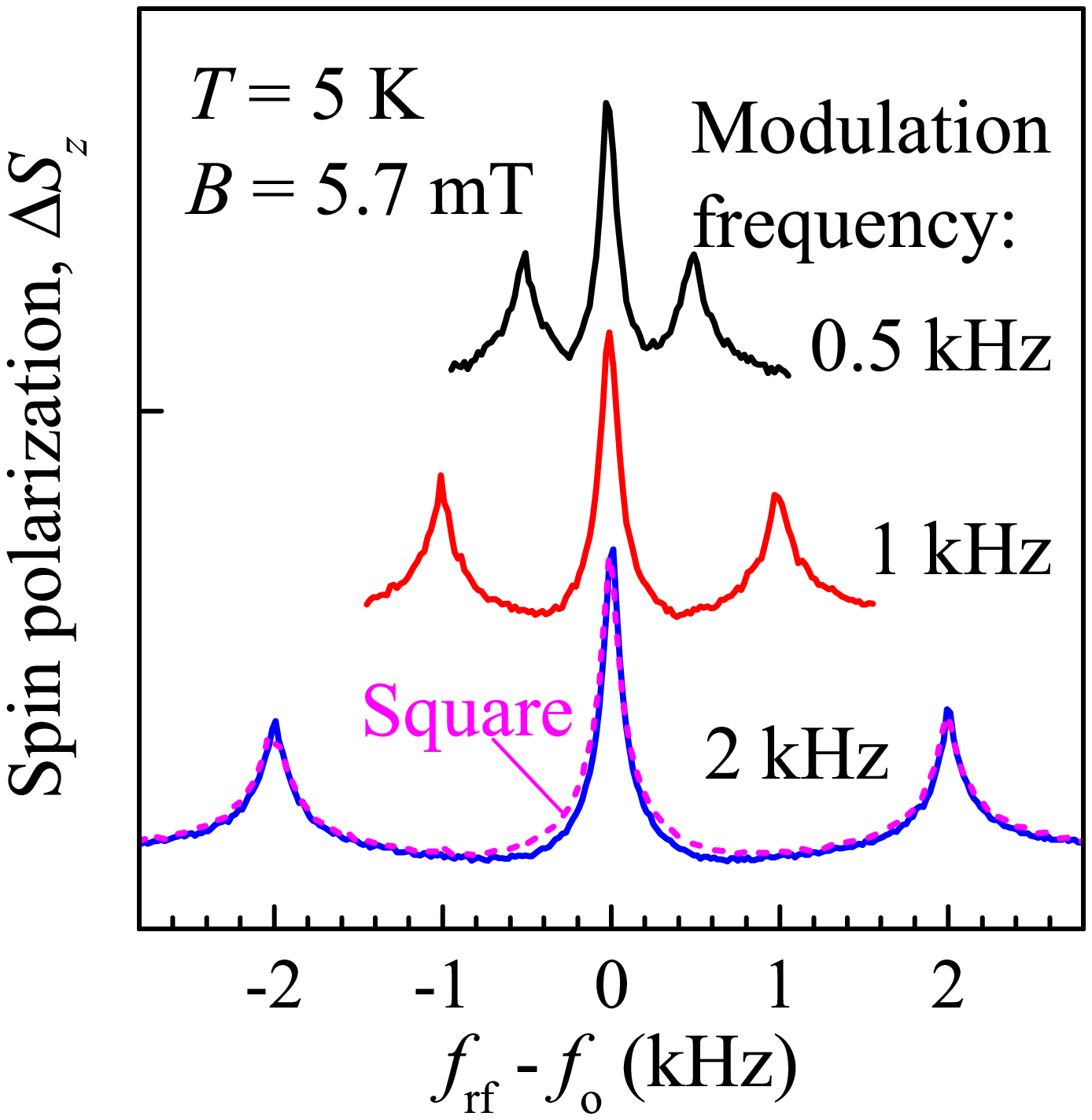}
\caption{SRSA spectra for the different modulation frequencies of the rf field. The modulation is done with the sinusoidal profile (solid lines) and with the square profile (dashed line). $B = 5.7$~mT, $P = 1$~mW, and $T = 5$~K.}
\label{fig:fmDep}
\end{figure}
\section{Effect of the rf field modulation}
In order to perform the synchronous detection revealing the effect of a rf field on the spin polarization, the rf field with the frequency $f_\text{rf} \approx 76$~MHz was modulated at a kilohertz frequency $f_\text{m}$, and at this frequency the lock-in amplifier was synchronized. In this way, the lock-in amplifier measures the signal difference with the high and low levels of the rf field amplitude. However, modulation also results in the additional modes of the oscillating rf field at the frequencies $f_\text{rf} \pm f_\text{m}$ and correspondingly in the additional SRSA peaks occurring at $f_\text{rf} = f_\text{o}\pm f_\text{m}$ (Fig.~\ref{fig:fmDep}). The shape of the SRSA spectrum is barely affected by the shape of the modulation profile. Figure~\ref{fig:fmDep} shows the exemplary comparison of the spectra for the sinusoidal (solid line) and square (dashed line) modulation profiles. The experiments presented in the main manuscript were done with the sinusoidal modulation at $f_\text{m} = 5$~kHz much higher the peak width, to avoid the interference from the side peaks.  

\section{Magnetic field dependence of SRSA spectrum at the increased laser power}
\begin{figure}[t]
\includegraphics[width=1\columnwidth]{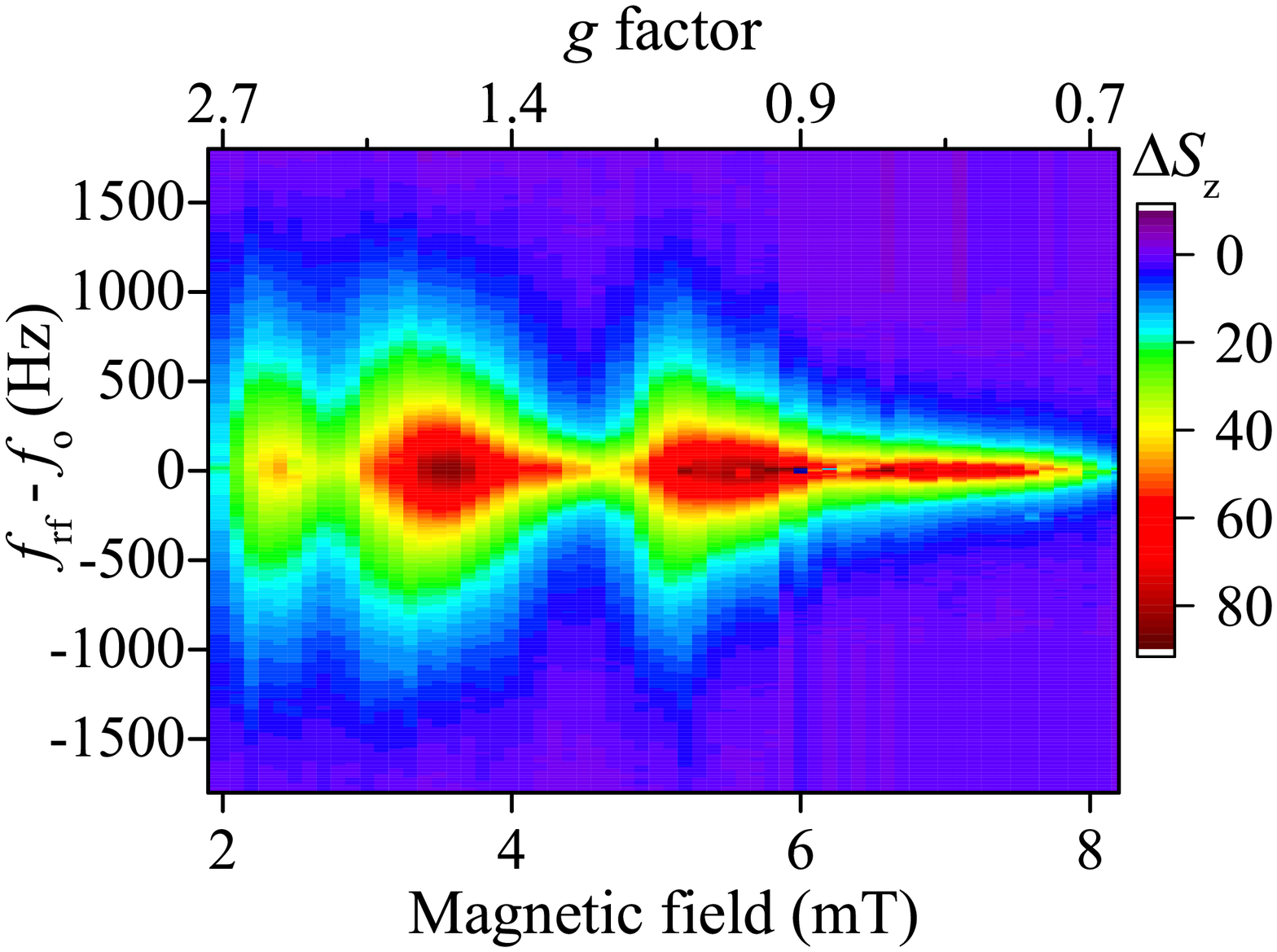}
\caption{Spin polarization as a function of the magnetic field (or the $g$ factor shown on the top axis) and the rf field frequency offset with respect to the laser pulse repetition frequency. $P = 1$~mW, and $T = 5$~K.}
\label{fig:BMap}
\end{figure}
Figure~\ref{fig:BMap} shows a color map representing the SRSA spectra (dependence of the spin polarization on the rf field frequency) at different magnetic fields for the increased laser power $P = 1$~mW (with respect to $P = 0.5$~mW used in the main text). The latter, despite the broadening of the SRSA spectra, results in the well resolved maxima at $B = 2.2, 3.3$, and $5.5$~mT corresponding to the $g$ factors $|g| = 2\pi\hbar f_\text{o}/\mu_\text{B} B = 2.5, 1.7$, and $1.0$.

\section{Dependence of SRSA spectrum on the rf field amplitude}

The amplitude of the rf field $b$ is controlled by the rf voltage $U_\text{rf}$ applied to the coil, so that $b = k U_\text{rf}$. The coefficient $k$ can be estimated as $k = 1/(2\pi^2 f_\text{rf} N r^2) \approx 0.13$~mT$/$V, where $r\approx 0.7$~mm is the coil radius, $N = 10$ is the number of windings \cite{Belykh2019}. The results presented above and in the main manuscript are obtained with $U_\text{rf} = 10$~V which correspond to $b = 1.3$~mT and Rabi frequency $\Omega_\text{R} = |g| \mu_\text{B} b /2\hbar \approx 2\pi \times 9$~MHz for $|g| = 1.0$. Comparable values of the rf magnetic field were achieved in Ref. \onlinecite{Belykh2020} where $b$ was determined from the model fit to the experimental data.

\begin{figure*}
\includegraphics[width=2\columnwidth]{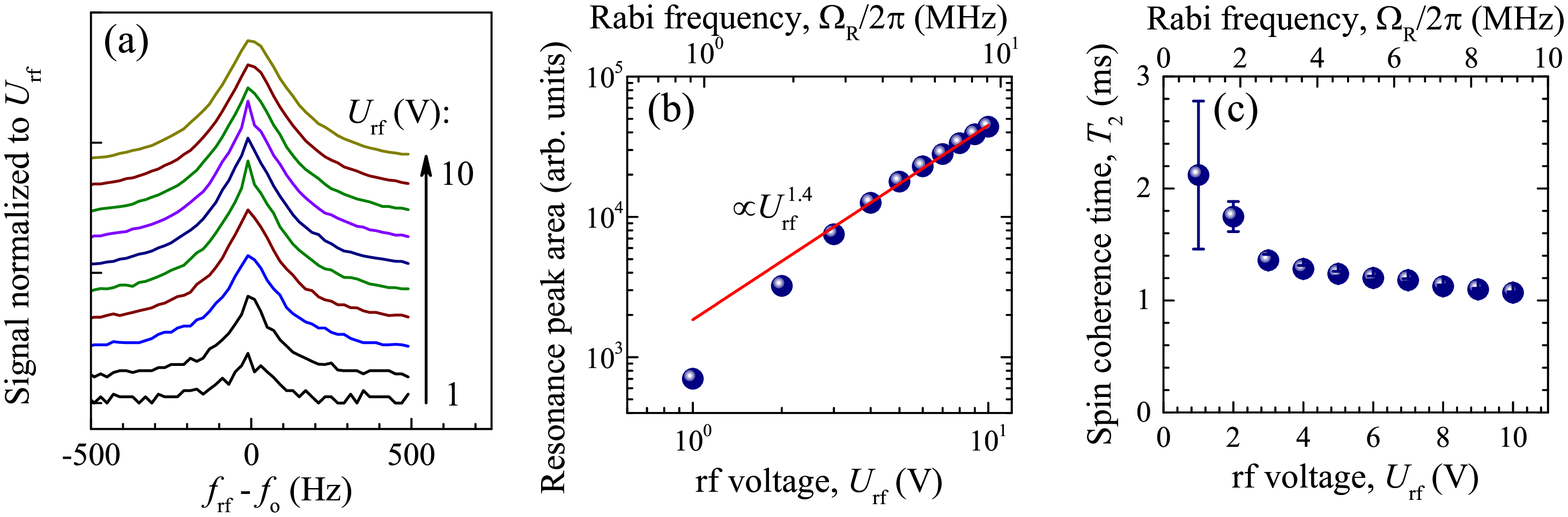}
\caption{(a) SRSA spectra for different values of the rf voltage $U_\text{rf}$. The spectra are normalized to $U_\text{rf}$.(b) Magnitude (area) of the SRSA resonance peak as a function of the rf voltage (bottom axis) or Rabi frequency (top axis). (c) Spin coherence time as a function of the rf voltage (bottom axis) or Rabi frequency (top axis). In (a)-(c) $B = 5.2$~mT, $P = 1$~mW, and $T = 5$~K.}
\label{fig:VDep}
\end{figure*}

Figure~\ref{fig:VDep}(a) shows the dependence of the SRSA spectra on the rf voltage. The magnitude (area) of the SRSA peak shows superlinear increase with $U_\text{rf}$ [Fig.~\ref{fig:VDep}(b)]. The increase is related to the involvement of more spins to the coherent precession at the frequency $f_\text{rf}$. Interesting, the spin coherence time $T_2$ shows decrease with $U_\text{rf}$ which can not be explained within our current understanding of the SRSA effect. This may be related to nonlinearities at the high rf fields, strong dependence of $T_2$ on the magnetic field  etc. 

Note that interesting physics related to the suppression of nuclear spin fluctuations by the rf field corresponds to the Rabi frequency $\Omega_\text{R}/2\pi$ lying in the kHz range when it is comparable to the inverse correlation time of nuclear fluctuations. In our experiments the minimal voltage at which the signal can be reliably registered corresponds to $\Omega_\text{R}/2\pi \sim 1$~MHz. One way to enter the above mentioned regime is to increase the magnetic field (and $f_\text{rf}$) by three orders of magnitude, which will decrease correspondingly the correlation time of nuclear spin fluctuations.   

\onecolumngrid
\twocolumngrid

\end{document}